\documentclass[twocolumn,pre,showpacs]{revtex4}
\usepackage{graphicx}
\usepackage{epsfig}
\usepackage{bm}

\begin{document}

\title{Quasispecies Theory for Multiple-Peak Fitness Landscapes}

\author{D.\ B.\ Saakian $^{1,2}$,
 E.\ Mu\~noz$^3$,
 Chin-Kun Hu$^{1}$, and
M.\ W.\ Deem$^3$
}

\affiliation{
\hbox{}$^1$Institute of Physics, Academia Sinica, Nankang,
Taipei 11529, Taiwan\\
\hbox{}$^2$Yerevan Physics Institute,
Alikhanian Brothers St. 2, Yerevan 375036, Armenia \\
\hbox{}$^3$Department of Physics \& Astronomy,
Rice University, Houston,Texas 77005--1892, USA}

\begin{abstract}
We use a path integral representation to
solve the Eigen and Crow-Kimura molecular evolution models
for the case of multiple fitness peaks with arbitrary
fitness and degradation functions.
%We present explicit results for the two peak case.
In the general case, we find that the solution to these molecular evolution
models can be written as the optimum of a fitness function, with
constraints enforced by Lagrange multipliers and with a term accounting
for the entropy of the spreading population in sequence space.
The results for the Eigen model are applied to consider virus or cancer
proliferation under the control of drugs or the immune system.
\end{abstract}


\maketitle

\section{Introduction}

It is widely accepted that
biological evolution proceeds on a rugged fitness landscape
\cite{wr31,ka87}.
In the last decade, several exact results \cite{bb97,sh04a,sh04b}
have been derived
for the molecular evolution models of Eigen
\cite{ei71,ei89} and Crow-Kimura \cite{ck70}
and their generalizations \cite{Lassig}.

In this paper, we consider a multiple-peak replication rate
landscape as a means to model a rugged fitness landscape.
To date, there are few rigorous results for multiple-peak fitness landscapes.
Such results begin to make the connection with the
biologically-relevant case of a rugged fitness landscape. We here derive
error thresholds by means of a path integral representation
for both the Eigen and Crow-Kimura
mutation-selection schemes with an arbitrary number of replication rate
peaks.

We first generalize the Crow-Kimura model to the multiple-peak case.
The solution of the single-peak version of this model,
where the replication rate is a function of the Hamming distance
from one configuration, was provided
by a path integral representation in \cite{sh04a,sh04b,sh04c}.
We here provide the solution to this model for $K$ peaks,
where the replication rate is a function of the Hamming distances
from $K$ configurations,
again  by means of
a path integral.  We find that the mean distances from the peaks
maximize the replication rate,
with constraints provided by Lagrange multipliers,
and with an additional term that represents the entropy of the
population in sequence space.  Explicit solutions to this maximization
task are given for the two-peak case.

We then generalize and
solve the continuous-time Eigen model for $K$ peaks,
where the replication and degradation rates
are functions of the Hamming distances from $K$ configurations,
A solution of the discrete-time, single-peak Eigen model, which
in a sense interpolates between the Crow-Kimura and continuous-time
Eigen model \cite{Krug}, was provided in \cite{Peliti2002}.
We here solve the continuous-time Eigen
model for $K$ peaks, again by means of a path integral representation.
The mean distances from the peaks maximize an excess
replication rate
with an effective mutation rate, with constraints provided by
Lagrange multipliers, and with an additional term that represents
the entropy of the population in sequence space under the effective
mutation rate.

The Eigen model was first developed to study viral evolution \cite{ei71},
and we use our solution of the two-peak Eigen model to consider viral
propagation in the presence of either immune system suppression
or an anti-viral drug.
The preferred viral genome exists at one point in genome space.
Conversely, the drug  or immune system suppresses the virus
most strongly at some other point in genome space.  These two
points in genome space are the two peaks of the model.  The viral
growth rate and the suppression rate both decrease with the Hamming
distance away from these two unique points.

The rest of the paper is organized as follows.
In Section 2  we describe the generalization of the Crow-Kimura, or
parallel, model to multiple peaks.  We provide a solution of this
model for an arbitrary replication rate function that depends on
distances from $K$ peaks.
In Section 3,  we describe the Eigen model.
We provide a solution for arbitrary replication and degradation rate
functions  that depend on distances from $K$ peaks.
In Section 4, we use the Eigen model theory to address
the interaction of the immune system with a drug.
We consider both adaptable drugs and the original antigenic
sin phenomenon \cite{de03}.  We also consider tumor suppression  by
the immune system.  We discuss these results and conclude in
section 5.  We provide a derivation of the path integral
representation of the continuous-time Eigen model in the Appendix.

\section{Crow-Kimura model with multiple peaks}
Here we first briefly introduce the Crow-Kimura model \cite{ck70}
and its quantum spin version \cite{bb97} so that it is easier to
understand its generalizations to be studied in the present paper.
In the Crow-Kimura model, any genotype configuration $i$ is
specified a sequence of $N$ two-valued spins $s_n=\pm 1$, $1
\le n \le N$. We denote such configuration $i$ by $S_i\equiv
({s_1^i, \dots, s_N^i})$.
That is, as in \cite{ei89},
we consider $s_n= +1$ to represent the purines (A, G) and
$s_n = -1$ to represent the pyrimidines (C, T).
The difference between two
configurations $S_i$ and $S_j \equiv ({s_1^j, \dots, s_N^j})$
is described by the Hamming distance
$d_{ij}=(N-\sum_n s_n^i s_n^j)/2$, which is the number of
different spins between $S_i$ and $S_j$. The relative frequency
$p_i$ of configurations with a given sequence $1 \le i \le 2^N$ satisfies
\begin{eqnarray}
\label{e1} \frac{{dp}_i}{dt}={p_i}(r_i-{\sum}_{j=1}^{2^N}r_j
p_j)+{\sum}_{j=1}^{2^N}\mu_{i j}p_j \ .
\label{e0}
\end{eqnarray}
Here $r_i$ is the the replication rate
 or the number of offspring per
unit period of time, the fitness, and $\mu_{ij}$ is the mutation rate to move
from sequence $S_i$ to sequence
$S_j$ per unit period of time.   In the Crow-Kimura model, only
single base mutations are allowed:
$\mu_{ij}=\gamma \Delta(d_{ij}-1) - N \gamma \Delta(d_{ij})$.
Here $\Delta(n)$ is the Kronecker delta.

The fitness of an organism with a given genotype is specified
in the Crow-Kimura model by the choice of
the replication rate function $r_i$, which is a function
of the genotype: $r_{i}=f(S_i)$. It has been observed
\cite{bb97} that the system (\ref{e0}),
 with $r_i\equiv f(s_1^i,\dots,s_N^i)$
evolves according to a Schr\"odinger equation in imaginary time with
the Hamiltonian:
\begin{equation}
\label{e2}
-H=\gamma\sum_{n=1}^N(\sigma^x_n-1)+f(\sigma_1^z, \ldots, \sigma_N^z) \ .
\end{equation}
Here $\sigma^x$ and $\sigma^z$
are the Pauli matrices.
The mean replication rate, or fitness,
of the equilibrium population
of genotypes is calculated as
$\lim_{t \to \infty} \sum_i p_i(t) r_i =
 \lim_{\beta \to \infty} \frac{1}{\beta}\ln Z\equiv
\lim_{\beta \to \infty} \frac{1}{\beta}
\ln Tr \exp(-\beta H)$ \cite{ei89}.
In this way it is possible to find the phase structure and error threshold
of the equilibrium population.  In the generalized setting, the
Crow-Kimura model is often called the parallel model.

\subsection{The parallel model with two peaks}
We consider two peaks to be located at
two configurations $v^1_n,v^2_n,1\le n\le N$,
where $v^1_n=\pm 1,v^2_n=\pm 1$, and the two configurations have $l$
common spins:
 $\sum_{n=1}^Nv^1_nv^2_n=2l-N$.
The value of $l$ determines how close the two peaks are in
genotype space.
Now the replication rate of configuration $S$ is a function of the
Hamming distances to each peak:
\begin{equation}
\label{e3} r_i=f(2L_1/N-1,2L_2/N-1) \ ,
\end{equation}
where $\sum_{n=1}^Nv^1_n s_n= 2 L_1-N$ and
$ \sum_{n=1}^Nv^2_n s_n= 2 L_2-N$.

Due to the symmetry of the Hamiltonian, the equilibrium
frequencies are a function only of the distances from
the two peaks:
 $p_i\equiv p(L_1,L_2)$. We
define the factors $x_{\alpha_1,\alpha_2}$ that describe the fraction
of spins a
configuration $S$ has in common with the spins of configurations $v^1, v^2$.
In particular, we define the fraction of spins that are equal
to $\alpha_k$ times the value in peak configuration $v^k$.  For $K$ peaks, the
general definition is
$x_{\alpha_1 \ldots \alpha_K}=(1/N) \sum_{n=1,N}\delta[s_n,\alpha_1 v_n^1]
\dots\delta[s_n,\alpha_K v_n^K]$.  For the two peak case, these
factors are related to the distances from the configuration to
each peak and to the distance between the peaks:
\begin{eqnarray}
\label{e4}
x_{+-}(L_1,L_2)&=&(L_1-L_2+N-l)/(2N)\nonumber\\
x_{++}(L_1,L_2)&=&(L_1+L_2-N+l)/(2N)\nonumber\\
x_{--}(L_1,L_2)&=&(-L_1-L_2+N+l)/(2N)\nonumber\\
x_{-+}(L_1,L_2)&=&(-L_1+L_2+N-l)/(2N) \ .
\end{eqnarray}
With these factors $x$,
we find the following equation for the total probability
at a given value of $L_1$ and $L_2$, $P(L_1,L_2)$:
\begin{eqnarray}
\label{e5}
&&\frac{dP(L_1,L_2)}{dt}= f(\frac{2L_1}{N}-1,\frac{2L_2}{N}-1) P(L_1,L_2)
 \nonumber\\ &&
-\gamma N P(L_1,L_2)
 \nonumber\\ &&
+\gamma  \sum_{\alpha_1=\pm 1,\alpha_2=\pm 1}
    N x_{\alpha_1,\alpha_2}(L_1+\alpha_1,L_2+\alpha_2)
 \nonumber\\ &&
 \times P(L_1+\alpha_1,L_2+\alpha_2)
\nonumber \\
&& -P(L_1,L_2)
\sum_{L_1',L_2'=0}^N
f(\frac{2L_1'}{N}-1,\frac{2L_2'}{N}-1)
 P(L'_{1},L'_{2}) \ .
\nonumber \\
\end{eqnarray}
Only the values of $L_1$ and $L_2$ satisfying
the conditions $0 \le L_i \le N$,
$\vert L_1 + L_2 - N \vert \le l$,
$\vert L_1 - L_2 \vert \le N-l$ are associated with
non-zero probabilities.
Equation (\ref{e5}) can be solved numerically to find the error threshold
and average distance of the population to the two peaks.  In the next section
we solve this equation, and its generalization to $K$ peaks, analytically.

\subsection{Exact solution of the $K$ peak case by a path integral representation}
We consider the case of $K$ peaks. We consider the replication
rate to depend only on the distances from each peak
\begin{eqnarray}
\label{e6} r_i=f(\frac{2L_1}{N}-1,\dots ,\frac{2L_K}{N}-1)\equiv
Nf_0(u_1,\ldots,u_K) \ ,
\end{eqnarray}
where $N u_k = \sum_{n=1}^N v^k_n s_n = 2 L_k-N, 1\le k\le K$. The
observable value $\langle u_k \rangle $
is called the surface magnetization \cite{Tarazona}, or
surplus \cite{bb97}, for peak $k$.

Characterization of the fitness function that depends on $K$ peaks
through the $K$ values of $u_k$ requires more than the 
$K(K-1)/2$ Hamming distances between the peaks. It proves convenient
to define the $2^K$ parameters $y_{\alpha_1 \ldots \alpha_K}
\equiv y_i, i \le i \le 2^K$. These are defined by $y_i =(1/N)
\sum_{n=1}^N \prod_{k=1}^K \delta(\alpha_{i k}, v_n^k)$. Here
$\alpha_i$ is the set of indices $\alpha_1 \ldots \alpha_K$, and
$\alpha_{ik} = \alpha_k$ in a $i$-th set of indices $\alpha_1
\ldots \alpha_K$. The introduction of the $2^K$ parameters $y_i$ is
one principle point of this article.

The Trotter-Suzuki
method has been applied in  \cite{sh04b,sh04c}
to convert the quantum partition function for a single peak model
into a classical functional
integral.
While
calculating $Z\equiv \exp[-\beta H]$,
intermediate spin
configurations are introduced. We find $Z$ is a
functional integral, with the integrand involving
a partition function
of a spin
system in 2-$d$ lattice. In the spin system,
there is a nearest-neighbor interaction
in horizontal direction and a mean-field like interaction in the vertical
direction.   This spin system
partition function was evaluated in \cite{sh04b,sh04c}
under the assumption that the field values are constant.
A path integral representation of the discrete time Eigen model,
which is quite similar to the parallel model, was introduced
by Peliti \cite{Peliti2002}.

We here generalize this procedure to $K$ peaks and calculate
the time-dependent path integral and Ising partition function.
Since the replication rate is a function of $K$ distances, the
functional integral is over
$K$ fields that represent the $K$ magnetizations. The path integral
form of the partition function is
\begin{eqnarray}
\label{e7d}
Z &=& \int {\cal D} M_k {\cal D} H_k \exp \bigg\{  N \int_0^\beta d \beta'
 \bigg[ f_0[M_1(\beta'),\ldots,M_K(\beta')]
\nonumber \\ &&
-\sum_{k=1}^K H_k(\beta') M_k(\beta') - \gamma \bigg]
 + N \sum_{i=1}^{2^K}  y_i  \ln Q_1 \bigg\}  ,
\end{eqnarray}
where
\begin{eqnarray}
\label{e7c}
Q_1 = Tr \hat T
e^{
\int_0^\beta d \beta'
  \left[\sigma^x \gamma + \sigma^z \sum_{k=1}^K \alpha_{ik} H_k(\beta')
\right] } \ .
\end{eqnarray}
Here $\beta = t$, is the large
time to which Eq.\ (\ref{e0}) is solved,
and the operator $\hat T$ denotes time ordering
\cite{sh04c}, discussed in the Appendix in the context of the Eigen
model.
Using that $N$ is large, we take the saddle point.
Considering $\delta \ln Z / \delta M_k(\beta') = 0$ and
$\delta \ln Z / \delta H_k(\beta')=0$, we find
$M_k(\beta')$ and $H_k(\beta')$ independent of $\beta'$ is a solution.
At long time,
therefore, the mean replication rate, or fitness, per site becomes
\begin{eqnarray}
\label{e7a}
\frac{\ln Z}{\beta N} &=& \max_{M_k,H_k}  \bigg\{
f_0(M_1,\ldots,M_K)-\sum_{k=1}^K H_k M_k - \gamma
\nonumber \\ &&
+ \sum_{i=1}^{2^K} y_i \left[\gamma^2 +(\sum_{k=1}^K\alpha_{ik}
H_k)^2 \right]^{1/2}  \bigg\} \ .
\end{eqnarray}
We take the saddle point in $H_k$ to find
\begin{eqnarray}
\label{e8}
M_k=
\sum_i y_{i}\alpha_{ik}\frac{\sum_{k'=1}^K\alpha_{i k'}H_{k'} }{\sqrt{\gamma^2
+(\sum_{k'=1}^K\alpha_{i k'}H_{k'})^2}} \ .
\end{eqnarray}
We note that the observable, surface, magnetization given by
 $\langle u_k \rangle $ is
not directly accessible in the saddle point limit, but is calculable from the
mean replication rate \cite{bb97}.
In the one peak case one defines the observable
surface magnetization for a monotonic  fitness function
as follows \cite{bb98}: one solves the equation
 $f_0(\langle u \rangle)= (\ln Z) / (\beta N)$.
For multiple peaks, we use this same trick, considering a
symmetric fitness function and assuming $\langle u_1 \rangle
= \langle u_2 \rangle \dots = \langle u_K \rangle$.

\subsection{Explicit results for the two peak case}
For clarity, we write the expression for the case of two peaks.
In this case,
$ y_{++} + y_{--} =(1+m)/2$ and $ y_{+-} + y_{-+} =(1-m)/2$,
where $m=(2l-N)/N$.  We solve Eq.\ (\ref{e8}) for
the fields $H_k$ and put the result into Eq.\ (\ref{e7a}).
We find that for a pure phase, the bulk magnetizations
maximize the function
\begin{eqnarray}
\label{e9a}
\frac{\ln Z}{N \beta} &=& f_0(M_1, M_2)
\nonumber \\ &&
+ \frac{\gamma}{2}\sqrt{(1+m)^2-(M_1+M_2)^2}
\nonumber \\ &&
+ \frac{\gamma}{2}\sqrt{(1-m)^2-(M_1-M_2)^2}  - \gamma \ ,
\end{eqnarray}
with the constraints
\begin{eqnarray}
\label{e10}
-1\le M_1\le 1; \quad-1\le M_2\le 1
\nonumber\\
-(1+m) \le M_1+M_2 \le 1+m
\nonumber\\
\quad -(1-m) \le M_1-M_2 \le 1-m \ .
\end{eqnarray}
In the case of a
quadratic replication rate, $f_0=k_1M_1^2+k_2M_2^2+k_3M_1M_2$,
Eq.\ (\ref{e9a}) becomes
\begin{eqnarray}
\label{e9}
\frac{\ln Z}{N \beta} &=& k_1 M_1^2+k_2 M_2^2+k_3 M_1 M_2  \nonumber\\
&&+ \frac{\gamma}{2}\sqrt{(1+m)^2-(M_1+M_2)^2}
\nonumber \\ && +
\frac{\gamma}{2}\sqrt{(1-m)^2-(M_1-M_2)^2}  - \gamma \ ,
\end{eqnarray}
with the constraints of Eq.\ (\ref{e10}).

As an example, we consider the replication rate function
$f_0=k(M_1^2+M_2^2+M_1 M_2)/2$. When $m \ge 0$, and the
two peaks are within a Hamming distance of $N/2$ of each other, there is
a solution with $M_1=M_2=M$ for which
\begin{eqnarray}
\label{e10c}
&& \frac{3 k M^2}{2\gamma}+\left[\left(\frac{1+m}{2}\right)^2
   -M^2\right]^{1/2} -\frac{1+m}{2}
\nonumber \\ && =
\frac{k}{2\gamma} \left(\langle u_1 \rangle ^2 + \langle u_2 \rangle^2 +
\langle  u_1 \rangle \langle u_2 \rangle \right) \ ,
\end{eqnarray}
where the observable, surface, magnetization
is given by $ \langle u_i \rangle = \langle 2
L_i/N - 1 \rangle$. We find
\begin{eqnarray}
\label{e10d}
M_1=M_2=M=\sqrt{(1+m)^2/4-\gamma^2/(9k^2)} \ .
\end{eqnarray}
We have for the mean replication rate, or fitness, per site
\begin{eqnarray}
\label{e10e}
\frac{\ln Z}{\beta N}
 =\frac{3k}{2} \left(\frac{1+m}{2}-\frac{\gamma}{3k} \right)^2 \ ,
\end{eqnarray}
so that  \cite{bb97}
\begin{eqnarray}
\label{e10f}
\langle u \rangle = \frac{1+m}{2}-\frac{\gamma}{3k} \ .
\end{eqnarray}
When $m < 0$, and the
two peaks are greater than a Hamming distance of $N/2$ of each other, 
 there is
a solution with $M_1=-M_2=M$ for which
\begin{eqnarray}
\label{e10g}
&&\frac{k M^2}{2\gamma}+\left[ (\frac{1-m}{2})^2-M^2 \right]^{1/2}
- \frac{1-m}{2} \nonumber \\ && =
  \frac{ k}{2 \gamma} \left(\langle u_1 \rangle^2 + \langle u_2 \rangle^2
 + \langle u_1 \rangle \langle u_2 \rangle \right) \ .
\end{eqnarray}
One solution is
\begin{eqnarray}
\label{e10h}
M_1=-M_2=\sqrt{(1-m)^2/4-\gamma^2/k^2} \ ,
\end{eqnarray}
which gives for a mean replication rate, or fitness, per site
\begin{eqnarray}
\label{e10i}
\frac{\ln Z}{\beta N} =\frac{k}{2}\left(\frac{1-m}{2}-\frac{\gamma}{k} \right)^2 \ ,
\end{eqnarray}
so that  \cite{bb97}
\begin{eqnarray}
\label{e10j}
\langle u_1\rangle = -\langle u_2\rangle = \frac{1-m}{2}-\frac{\gamma}{k} \ .
\end{eqnarray}
Numerical solution is in agreement with our analytical formulas, as
shown in Table \ref{tab1}.
\begin{table}[t!]
\begin{tabular}{c|c|c|c|c|c}
\em m &\em k &\em $\langle u_{1} \rangle$ &\em $\langle u_{2}
\rangle $
&\em $\langle u_1 \rangle_{\rm analytic} $
 &\em $\langle u_2 \rangle_{\rm analytic}$ \\\hline \hline
0.93 & 3.0 & 0.85 & 0.85 & 0.853 & 0.853  \\ \hline
0.93 & 2.0 & 0.80 & 0.80 & 0.798 & 0.798  \\ \hline
0.7  & 3.0 & 0.74 & 0.74 & 0.738 & 0.738  \\ \hline
0.7  & 2.0 & 0.68 & 0.68 & 0.683 & 0.683  \\ \hline
-0.7  & 3.0 & 0.52 & -0.52 & 0.517 & -0.517  \\ \hline
-0.7  & 2.0 & 0.35 & -0.35 & 0.35 & -0.35 \\ \hline
-0.93 & 3.0 & 0.63 & -0.63 & 0.631 & -0.631 \\ \hline
-0.93 & 2.0 & 0.46 & -0.46 & 0.465 & -0.465
\end{tabular}
\caption{Comparison between the analytical formulas Eqs.\
(\ref{e10f}, \ref{e10j}) for the
two peak landscape in the parallel model and
results from a
direct numerical solution of the system of differential equations,
Eq.\ (\ref{e5}), for sequences of length $N=1000$, with
$p(L_1, L_2, t=0) = \delta(L_1,N) \delta(L_2,l)$.
\label{tab1}
}
\end{table}

\section{Eigen model with multiple peaks}
\subsection{Exact solution by a path integral representation}
In the case of the Eigen model, the system is defined by means of
replication rate functions, $r_j$,
as well as degradation rates, $D_j$:
\begin{equation}
\label{e11} \frac{dp_i}{dt}=
\sum_{j=1}^{2^N}\left[Q_{ij}r_j-\delta_{ij}D_j \right] p_j-p_i \left[
\sum_{j=1}^{2^N}(r_{j}-D_j)p_j \right] \ .
\end{equation}
Here the frequencies of a given genome, $p_i$, satisfy $\sum_{i=1}^{2^N}p_i=1$.
The transition rates are given by
$Q_{ij}=q^{N-d(i,j)}(1-q)^{d(i,j)}$, with the distance between
two genomes $S_i$ and $S_j$ given by $ d(i,j)=
(N-\sum_{n=1}^N s_n^i s_n^j )/2$.  The parameter $\gamma=N(1-q)$ describes the
efficiency of mutations.  We take $\gamma = O(1)$.
As in Eq.\ (\ref{e6}), we take the replication and
degradation rate to depend only on the spin state, in particular
on the distances from each peak:
$r_i = f(S_i)$ and $D_i = D(S_i)$ where
\begin{equation}
 \label{e12}
f(S)=  Nf_0(u_1, \ldots, u_k),~D(S)= Nd_0(u_1, \ldots, u_k) \ .
 \end{equation}
We find the
path integral representation of the partition function for the
Eigen model for the $K$ peak case in the limit of long time as
 \begin{eqnarray}
\label{e13}
Z&=&
\int {\cal D} M_k
{\cal D} H_k
{\cal D} m_0
{\cal D} h_0
 \exp\bigg\{
N \int_0^\beta d \beta'  \bigg[
 \nonumber\\ &&
f_0(M_1, \ldots, M_K)e^{-\gamma(1-m_0)}- h_0 m_0
 \nonumber\\ &&
-\sum_{k=1}^K H_k M_k -d_0(M_1, \ldots, M_K) \bigg]
 \nonumber\\ &&
+ N
\sum_{i=1}^{2^K} y_{i} \ln Q_1
\bigg\} \ ,
\end{eqnarray}
where
\begin{eqnarray}
\label{e13a}
Q_1 = Tr \hat T
e^{
\int_0^\beta d \beta'
  \left[\sigma^x h_0(\beta') + \sigma^z \sum_{k=1}^K \alpha_{ik} H_k(\beta')
\right] } \ .
\end{eqnarray}
The $M_k$ are the values of the magnetization, and $\gamma m_0$ is an
effective mutation rate.
This form is derived in the Appendix.
Using that $N$ is large, we take the saddle point.
As before,
we find the mean excess replication rate per site, $f_m = \lim_{t \to \infty}
 \sum_i p_i(t) (r_i - D_i)/N$,
from the maximum of
the expression for $Z={\rm Tr} \exp(-\beta H)$.  We find
$Z \sim  \exp(\beta N f_{m})$, where
 \begin{eqnarray}
\label{e14}
f_{m}&=&f_0(M_1, \ldots, M_K)e^{-\gamma(1
-\sum_{i=1}^{2^K}y_i\sqrt{1-m_i^2})}
\nonumber \\ &&
-d_0(M_1, \ldots, M_K) \ ,
\end{eqnarray}
and where $m_0, M_k$ are defined through the fields $H_k$:
\begin{eqnarray}
\label{e15}
M_k &=& \sum_i y_i \alpha_{ik}
\frac{\sum_{k'=1}^K \alpha_{ik'} H_{k'}}
{\sqrt{h_0^2+(\sum_{k'=1}^K \alpha_{ik'} H_{k'})^2}}\nonumber\\
m_0 &=& \sum_i y_i
\frac{h_0} {\sqrt{h_0^2+(\sum_{k=1}^K \alpha_{ik} H_k)^2}} \ .
\end{eqnarray}
We define
\begin{eqnarray}
\label{e15c}
m_i=\frac{\sum_{k=1}^K \alpha_{ik} H_{k}}{\sqrt{h_0^2
+(\sum_{k=1}^K\alpha_{ik}H_{k})^2}} \ .
\end{eqnarray}
We thus find $m_0 = \sum_{i=1}^{2^K} y_i \sqrt{1-m_i^2}$, giving
Eq.\ (\ref{e14}).

\subsection{Eigen model with quadratic replication rate without degradation}
We apply our results to model
qualitatively the interaction of a virus
with a drug.
In some situations, one can describe the action of a drug against the
virus simply as a one peak Eigen model: that is, replication rate
is a function of the Hamming distance from one peak.
The virus may increase its mutation rate, and at some mutation rate there is
an error catastrophe \cite{ei02}.
Let us define the critical $\gamma$ for the replication rate function
\begin{eqnarray}
\label{e16}
f_0(M)=kM^2/2+1 \ .
\end{eqnarray}
According to our analytical solution, Eq.\ (\ref{e14}),
we consider the maximum of the mean excess replication rate per
site expression
\begin{eqnarray}
\label{e17}
f_0(M)\exp[-\gamma(1-\sqrt{1-M^2})] \ .
\end{eqnarray}
The error catastrophe occurs and leads to a phase
with $M=0$ when $ k < \gamma$.
The error threshold for this quadratic case is the same as in the case
of the Crow-Kimura model Eq.\ (\ref{e1}).
Numerical solution is in agreement with our analytical formulas, as
shown in Table \ref{tab2}.
\begin{table}[t!]
\begin{tabular}{c|c|c|c}
\em $k/\gamma$  &\em M &\em $\langle u \rangle$ &
\em $\langle u \rangle_{\rm analytic} $ \\
\hline \hline
1.2 & 0.24 & 0.065 & 0.068   \\ \hline
1.4 & 0.31 & 0.112 & 0.113 \\ \hline
1.6 & 0.35 & 0.146 & 0.147    \\ \hline
1.8 & 0.38 & 0.172 & 0.172     \\ \hline
2.0  & 0.41& 0.192 & 0.193   \\ \hline
\end{tabular}
\caption{Comparison between the analytical formulas Eqs.\
(\ref{e17}) for the 
quadratic landscape (\ref{e16}) in the Eigen model and
results from a
direct numerical solution of the system of differential equations \cite{ei89},
for sequences of length $N=4000$, with
$p(u, t=0) = \delta(u,1)$.
 We set $\gamma=5$.
\label{tab2}
}
\end{table}

\subsection{Simple formulas for two peak case}
In the two peak, $K=2$, 
case we can define  the $m_i$ from Eq.\ (\ref{e15c})
from the system
\begin{eqnarray}
M_1&=&\frac{1+m}{2}(m_1+m_2)+\frac{1-m}{2}(m_1-m_2)
\nonumber \\
M_2&=&\frac{1+m}{2}(m_1+m_2)-\frac{1-m}{2}(m_1-m_2) \ ,
\label{e19a}
\end{eqnarray}
where we have defined $m_1, m_2$ in terms of the $m_i$ from Eq.\ (\ref{e15c})
by
$m_{++} = -m_{--} = m_1+m_2$ and
$m_{+-} = -m_{-+} = m_1-m_2$ \ ,
We have for the mean excess replication rate per site
 \begin{eqnarray}
f_{m}&=&f_0(M_1,M_2)
\exp\bigg[-\gamma\bigg(1
\nonumber \\ && -\frac{1+m}{2}\sqrt{1 - (M_1+M_2)^2/(1+m)^2}
 \nonumber\\ &&
-\frac{1-m}{2}\sqrt{1-(M_1-M_2)^2/(1-m)^2}\bigg)\bigg]
 \nonumber\\ &&
-d_0(M_1,M_2) \ .
\label{e19}
\end{eqnarray}

\section{Biological applications}

The Eigen model is commonly used to consider virus or cancer
evolution.
We here consider an evolving virus or cancer and its control
by a drug or the immune system, using the $K=2$, two-peak version of
the Eigen model.  To model this situation, we consider there
to be an optimal genome for virus replication, and we
consider the replication rate function $f_0(M_1, M_2)$ to
depend only on the Hamming distance of the virus or cancer from
this preferred genome, $N (1-M_1)/2$.  Conversely, there is another
point in genome space that the drug or immune suppresses
most strongly, and we consider
the degradation rate function $d_0(M_1, M_2)$ to depend only on 
the Hamming distance from this point,
$N(1-M_2)/2$.  
While each of the functions $f_0$ and $d_0$ depend only
on one of the two distances, this is multiple-peak problem, because
both distances are needed to describe the evolution of
the system.

\subsection{Interaction of virus with a drug}
We first consider a virus interacting with a drug.  We
model this situation by the Eigen model
with one peak in the replication rate function and one
peak in the degradation rate function.
The virus replicates most quickly at one point in genome
space, with the rate at all other points given by
a function that depends on the Hamming distance from
this one point.
That is, in Eq.\ (\ref{e19}) we have
\begin{equation}
\label{e20a}
f_0(M_1, M_2)=\left\{
\begin{array}{lll}
 & A, & \,\,M_1 = 1 \\
 && \\
 & 1, &\,\,  M_1 < 1
\end{array} \right. \ .
\end{equation}
At another point in genome space, a drug suppresses the
virus most strongly.
We consider the case of exponential degradation,
a generic and prototypical example of recognition \cite{de03}:
\begin{equation}
\label{e20}
d_0(M_1, M_2)=e^{-b(1-M_2)} \ .
\end{equation}

\begin{figure}[t!]
\psfig{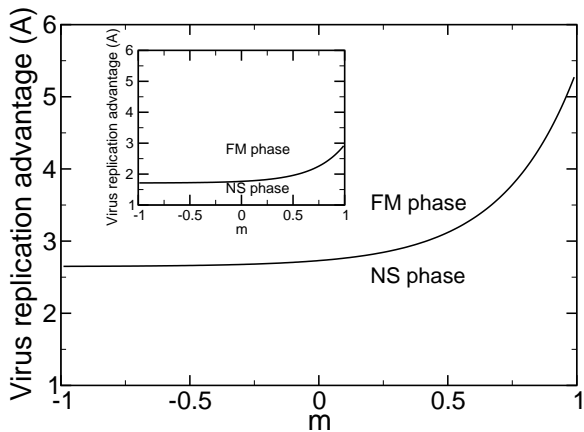}
\caption{A phase diagram for the interaction of virus and drug, according to
the narrow replication advantage model, Eqs.\ (\ref{e20}) and (\ref{e20a}).
We set
$b=3.5$ in the exponential degradation function Eq.\ (\ref{e20}), and
$\gamma = 1$.  
As the point in genome space
at which the drug is most effective moves toward the point
in genome space at which the virus grows most rapidly,
$m \to 1$,
a higher replication rate of the virus is required for the virus to
survive.
In the NS phase, the drug eliminates the virus.
In inset is shown
the phase diagram for the interaction of an adaptable
virus and a drug, according to
the flat peak replication advantage model, Eqs.\ (\ref{e20}) and (\ref{e24}).
We use
$M_0 = 0.9$ to represent a broad
peak for the virus replication rate.
\label{fig1}
}
\end{figure}

Applying the multiple-peak formalism, we find 
two phases.  There is a selected, ferromagnetic (FM) phase
 with $M_1=1$, $M_2=m$ and mean excess replication rate per site
\begin{eqnarray}
\label{e21}
f_{m}=Ae^{-\gamma}-\exp(-b(1-m)) \ .
\end{eqnarray}
There is also a non-selective $NS$ phase, with $M_1 < 1$.  The values of
$M_1$ and $M_2$ in the NS phase are those which maximize Eq.\ \ref{e19}
given the constraints of Eq.\ (\ref{e10}).
The error threshold corresponds to the situation when
the mean excess replication rate
of the FM and NS phases are equal.  The phase diagram
as a function of the optimal replication rate of the virus and
the distance between the points of optimal virus growth
and optimal virus suppression is shown in Fig.\ \ref{fig1}.
The optimal replication rate is $A$, and 
the distance between the points of optimal virus growth
and optimal virus suppression  is $N-l$, where the parameter
$m$ is defined as $m = (2l-N)/N$.
As the point in genome space
at which the drug is most effective moves toward the point
in genome space at which the virus grows most rapidly, 
the virus is more readily eradicated.  Alternatively, one can say
that
as the point in genome space
at which the drug is most effective moves toward the point
in genome space at which the virus grows most rapidly, 
a higher replication rate of the virus is required for its
survival.

\subsection{Interaction of an adaptable virus with a drug}

We now consider a virus that replicates with rate $A$ when the
genome is within a given Hamming distance from the optimal
genome and with rate $1$ otherwise.  That is,
in Eq.\ (\ref{e19}) we have
\begin{eqnarray}
 \label{e24}
f_0(M_1, M_2)=\left\{
\begin{array}{lll}
 & A, & \,\,M_0 \leq M_1 \leq 1 \\
 && \\
 & 1, &\,\,  -1 \leq M_1 < M_0
\end{array} \right. \ ,
\end{eqnarray}
where $M_0>0$ and close to one.  We suppression of the
virus by the drug as expressed
in Eq.\ (\ref{e20}).

There is again a ferromagnetic (FM) phase with a successful selection.
In the FM phase, one has $M_0 \le M_1 \le 1$.  The evolved values of
$M_1$ and $M_2$ maximize
 \begin{eqnarray}
\label{e25}
f_{m}&=&A
\exp\bigg[-\gamma\bigg(1
\nonumber \\ &&
-\frac{1+m}{2}\sqrt{1-(M_1+M_2)^2/(1+m)^2}\nonumber\\
&&-\frac{1-m}{2}\sqrt{1-(M_1-M_2)^2/(1-m)^2} \bigg) \bigg]
\nonumber \\ && -d_0(M_1,M_2) \ .
\end{eqnarray}
There is also a NS phase where the virus has been driven off
its advantaged peak, $M_1 < M_0$.
In this case, one seeks a maximum of Eq.\ (\ref{e19})
with $f_0 = 1$ via $M_1$ and $M_2$  in the range
$-1 \le M_1 \le M_0$, $-1 \le M_2 \le 1$, subject
to the constraints of  Eq.\ (\ref{e10}).

A phase diagram for this case is shown in the inset to
Fig.\ \ref{fig1}.  The broader
range of the virus fitness
 landscape allows the virus to survive
under a greater drug pressure in model Eq.\ (\ref{e24}) versus
Eq.\ (\ref{e20a}).  That is, as the point in genome space
at which the drug is most effective moves toward the point
in genome space at which the virus grows most rapidly, the adaptable
virus is still able to persist due to the
greater range of genotype space available in the FM phase.
For such an adaptable virus, a more specific, multi-drug cocktail
might be required for eradication.  A multi-drug cocktail
provides more suppression in a broader range of genome space, so that
the adaptable drug may be eradicated under a broader
range of conditions.  
%Thus, we have suggested an analysis of
%the fundamental evolutionary reasons for which a multi-drug
%cocktail is needed for adaptable viral diseases such as HIV,
%a result first deduced by a differential equation analysis of clinical
%immunology data \cite{Perelson}.

\subsection{Original antigenic sin}
\begin{figure}[t!]
\psfig{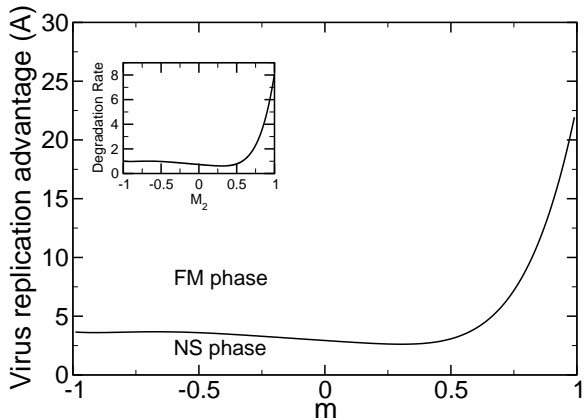}
\caption{A phase diagram corresponding to the original antigenic sin model
is shown. The degradation function (shown in inset) is chosen to closely
reproduce the binding constant behavior in \cite{de03}, with a limiting
degradation rate of $d_0(M_1, M_2 = -1) = 1$.
We use $\gamma=1$.
The virus replication advantage
required to escape immune system control is a non-monotonic
function of the Hamming distance,
$N (m-1)/2$,
 between 
the point in genome space
at which the immune system is most effective and
the point in genome space at which the virus grows most rapidly.
\label{fig3}
}
\end{figure}

The immune system acts much like a drug, as a natural protection
against death by infection.  Prior exposure such as vaccination
typically increases the immune control of a virus.
In some cases, the immune control
of a virus is non-monotonic in the distance between the vaccine
and the virus \cite{de03}.  This phenomenon is termed
original antigenic sin.  To model original antigenic sin, we consider
a non-monotonic degradation function, centered around the second peak,
which represents the non-monotonic behavior of the binding constant,
as in our previous model \cite{de03}.  We fit the binding
constant data \cite{de03} to a sixth order polynomial in $p$, where
$p = (1-M_2)/2$ is the relative distance between the recognition of
the antibody and the virus.  The degradation function is shown in inset
in Fig.\ \ref{fig3}.
We consider a single peak virus replication rate, Eq.\ (\ref{e20a}).

There is an interesting phase structure as a function of $m$.
From Eq.\ (\ref{e19}),
we have a FM phase with $M_1=1$, $M_2=m$.
We also have a non-selective NS phase, with $M_1 < 1$, where
$M_1, M_2$ are determined by maximization of  Eq.\  (\ref{e19})
with $f_0 = 1$ under the constraints of Eq.\  (\ref{e10}).
The phase diagram for typical parameters \cite{de03},
is shown in Fig.\ \ref{fig3}.
A continuous phase boundary is observed between the FM and
NS phases.  The virus replication rate required to escape eradication
by the adaptive immune system depends on how similar the
virus and the vaccine are.  When the vaccine is similar
to the virus, $m$ near 1, a 
large virus replication rate is required to escape eradication.
This result indicates the typical usefulness of vaccines in protection
against and eradication of viruses. When the vaccine is
not similar to the virus, $m < 0$, the vaccine is not effective, and
only a typical virus replication rate is required.

When the vaccine is somewhat similar to, but not identical to, the virus,
the replication rate required for virus survival is non-monotonic.  This
result is due
to the non-monotonic degradation rate around the vaccine degradation
peak.  The minimum in the required virus replication rate,
$m \approx 0.30$, corresponds to the
minimum in the degradation rate, $M_2 \approx 0.30$.
The competition between the immune system, vaccine, and virus
results in a non-trivial phase transition for the eradication
of the virus.

\subsection{Tumor control and proliferation}

We consider cancer to be a
mutating, replicating object, with a flat replication rate
around the first  peak, Eq.\ (\ref{e24}) and Eq.\ (\ref{e19}).
We consider the immune system to be able to eradicate the cancer
when the cancer is sufficiently different from self.  Thus, the
T cells have a constant degradation rate everywhere except near the
self, represented by the second peak:
\begin{eqnarray}
d_0(M_1, M_2)=\left\{
\begin{array}{lll}
 & B, & \,\, -1 \leq M_2 < M_b  \\
 && \\
 & 0, &\,\,  M_b \leq M_2 \leq 1
\end{array} \right. \ .
\end{eqnarray}
To be consistent with the biology,
we assume $M_b > 0$.  We also assume $M_0 > 1/2$.
Typically, also, the Hamming distance between
the cancer and the self will be small, $m$ will be positive and
near unity, although we do not assume this.

There are four possible selective, ferromagnetic phases.
We find the phase boundaries analytically, as a function of
$m = (2l - N)/N$.
For $ m M_0 < M_b$,
there is a FM4 phase with
$M_1 = M_0$ and $M_2 = m M_0$.
The mean excess replication rate per site
is $f_m = A e^{\gamma(\sqrt{1-M_0^2}-1)} - B$.
There is a FM3 phase with $M_1 = M_0$ and $M_2 = M_b$.
The mean excess replication rate per site is
$f_m = A e^{\gamma( \sqrt{(1+m)^2 - (M_0 + M_b)^2}
+ \sqrt{(1-m)^2 - (M_0 - M_b)^2} - 2)/2}$.
This phase is chosen over the FM4 phase when the mean
excess replication rate is greater.
For $m M_0 \ge M_b$ there is a FM2 phase with
$M_1 = M_0$ and $M_2 = m M_0$.
The mean excess replication rate per site
is $f_m = A e^{\gamma(\sqrt{1-M_0^2}-1)}$.
For $m M_b \ge M_0$ there is a FM1 phase with
$M_1 = m M_b$ and $M_2 = M_b$.
The mean excess replication rate per site
is $f_m = A e^{\gamma(\sqrt{1-M_b^2}-1)}$.

There are two non-selective phases.  There is a
NS1 phase  with
$M_1 = m M_b$ and $M_2 = M_b$.
The mean excess replication rate per site
is $f_m = e^{\gamma(\sqrt{1-M_b^2}-1)}$.
%For $m \ge M_b$ there is a NS2 phase  with
%$M_1 = M_b/m$ and $M_2 = M_b$.
%The mean excess replication rate per site
%is $f_m = e^{\gamma(\sqrt{1-M_b^2/m^2}-1)}$.
There is a NS2 phase with
$M_1 = M_2 = 0$.
The mean excess replication rate per site
is $f_m = 1-B$.

\begin{figure}[t!]
\psfig{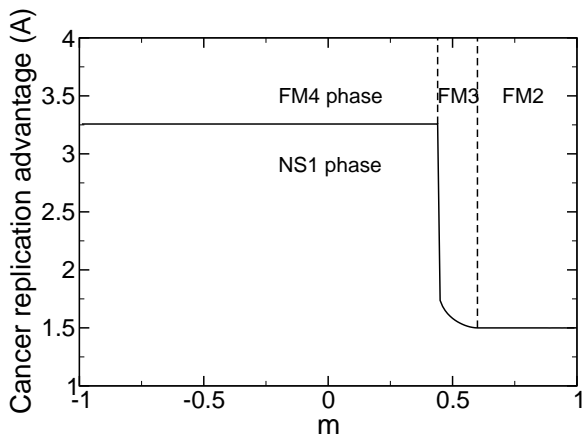}
\caption{A phase diagram corresponding to the immune control of
cancer
is shown.
We use $\gamma=1$, $B = 1$, $M_0 = 0.9$, and $M_b = 0.54$.
The cancer replication advantage
required to escape immune system control decreases
as the Hamming distance between the cancer and the self
decreases, 
$m \to 1$.
Three of the four selective and one of the two
non-selective phases are present for the
chosen parameters.
\label{fig4}
}
\end{figure}

In Fig.\  \ref{fig4} is shown the phase diagram for cancer proliferation.
According to our previous model \cite{de03,de04}, we choose
$(1-M_b)/2=0.23$.  We choose $M_0=0.9$
for the width of the advantaged cancer phase.  We choose the
immune suppression rate as $B = 1$.
As the cancer becomes more similar to the self, the immune control
becomes less effective, and the replication rate required
for the cancer to proliferate becomes less.  Three of
the four selective and one of the two non-selective phases
are present for this set of parameters.

\section{Discussion and Conclusion}

We have used the Eigen model to consider the interaction of
a virus or cancer with a drug or the immune system.  One can
also use the parallel
model to represent the replication dynamics of
the virus or the cancer.  This would  be an interesting application
of our formalism.

Another application of the formalism
would be to consider explicitly the degradation induced by
multi-drug cocktails.  That is, one would consider one peak to represent
the preferred virus genome and $K-1$ degradation peaks to represent
the $K-1$ drugs.  We note that in the general case, the $y_i$ parameters
depend on the explicit location of the drug degradation peaks, not
simply the distance between them.  Results from this application
of the formalism could be quite illuminating as regards the evolution
of multi-drug resistance.

In conclusion, we have solved two common
evolution models with general fitness, or replication and degradation rate,
landscapes that
depend on the Hamming distances from several fitness peaks.
Why is this important?  First, we have solved the microscopic
models, rather than assuming a phenomenological macroscopic model.
As is known in statistical mechanics, a phenomenological model
may not always detect the fine structure of critical
phenomena.  Second, approximate or numerical solutions, while
useful, do not always explicitly demonstrate the essence of the phenomenon.
With analytical solutions, the essence of
phenomenon is transparent.  Third, we have derived the first path
integral formulation of the Eigen model.  This formulation
may prove useful in other studies of this model of molecular evolution.

Our results for cancer are a case in point.  There are four
stable selective phases and two stable non-selective phases.
These results may help to shed light on the, at present,
poorly understood phenomena of interaction with the immune system, and
on why the immune response to cancer and to viruses differs in
important ways.  These phases could well also be related to the
different stages, or grades, through which tumors typically progress.

Our results are a first step toward making the connection
with evolution on rugged fitness landscapes, landscapes widely accepted to
be accurate depictions of nature.
We have
applied our solution of these microscopic complex adaptive systems to
model four situations in biology:
how a virus interacts with drug, how an adaptable virus interacts
with a drug, the problem of original antigenic
sin \cite{de03}, and immune system control of a proliferating
cancer.

\section*{Acknowledgments}  This work has been supported
by the following grants:
CRDF ARP2--2647--YE--05,
NSC 93--2112--M001--027, 94--2811--M001--014, AS--92-TP-A09, and
DARPA \#HR00110510057.

\section*{Appendix}

In this section, we derive the path integral representation for the
solution to the Eigen model.  For simplicity, we show the
derivation for the $K=1$ case. To our knowledge, this is the
first path integral expression representation of the solution
to the Eigen model.
This path integral representation allows us to
make strong analytic progress.  We start from the quantum
representation of the Eigen model \cite{sh04a,sh04c}.  The
Hamiltonian is given by
\begin{eqnarray}
-H &=& \sum_{l=0}^N N e^{-\gamma} \left( \frac{\gamma}{N} \right)^l
\sum_{1 \le i_1 < i_2 < \ldots <i_l \le N}
\sigma_{i_1}^x
\sigma_{i_2}^x \ldots
\sigma_{i_l}^x
\nonumber \\ &&
\times
f_0(\sigma^z) - N d_0(\sigma^z)
\nonumber \\
&\approx& N e^{-\gamma} e^{(\gamma/N) \sum_i \sigma_i^x} f_0(\sigma^z) -
 N d_0(\sigma^z) \ ,
\label{a1}
\end{eqnarray}
where we have used the fact that with $\gamma/N$ small, we need to
consider only $l \ll N$ spin flips.
The partition function is decomposed by a Trotter factorization:
\begin{eqnarray}
Z &=& Tr e^{-\beta H}
\nonumber \\ &=&
Tr
\langle S_1 \vert e^{-\beta H/L} \vert S_2 \rangle
\langle S_2 \vert e^{-\beta H/L} \vert S_3 \rangle
\ldots
\nonumber \\ && \times
\langle S_L \vert e^{-\beta H/L} \vert S_1 \rangle \ .
\label{a2}
\end{eqnarray}
Here
\begin{eqnarray}
&& \langle S_{l-1} \vert e^{-\beta H/L} \vert S_l \rangle
\nonumber \\
&=&\langle S_{l-1} \vert e^{
(\beta N/L) \bigg[
 e^{-\gamma} e^{(\gamma/N) \sum_i \sigma_i^x}
f_0(\sigma^z)
-d_0(\sigma^z)  \bigg]
}
\vert S_l \rangle
\nonumber \\ &=&
\langle S_{l-1}  \vert
I + \frac{\beta N}{L}  \bigg[
e^{-\gamma} e^{(\gamma/N) \sum_i \sigma_i^x}
f_0(\sum_n s_n^l/N)
\nonumber \\ && -
d_0(\sum_n s_n^l/N)  \bigg]
\vert S_l \rangle \ .
\label{a3}
\end{eqnarray}
We use the
notation $M_l = \sum_n s_n^l/N$.  We find
\begin{eqnarray}
\alpha_l & =& \langle S_{l-1} \vert
I + \frac{\beta N}{L}
e^{-\gamma} f_0(\sum_n s_n^l/N)
e^{(\gamma/N) \sum_i \sigma_i^x}
\nonumber \\ &&
 - \frac{\beta N}{L}  d_0(\sum_n s_n^l/N)
\vert S_l \rangle
\nonumber \\ &=&
\langle S_{l-1} \vert S_l \rangle
\left[1 - \frac{\beta N}{L}  d_0(M_l) \right] 
\nonumber \\ &&
+\frac{\beta N e^{-\gamma}}{L} f_0(M_l)
e^{B \sum_n (s^{l-1}_n s^l_n -1)} \ ,
\label{a5}
\end{eqnarray}
where $e^{-2 B} = \gamma / N $.
We thus find
\begin{eqnarray}
\alpha_l = \Delta(d_l)
\left[1 - \frac{\beta N}{L}  d_0(M_l) \right]
+ \frac{\beta N}{L}
e^{-\gamma} f_0(M_l) e^{B d_l} \ ,
\label{a6}
\end{eqnarray}
where $d_l = \sum_n (s^{l-1}_n s^l_n -1)$.
To represent this in path integral form, we consider
\begin{eqnarray}
 &&\frac{1}{(2 \pi)^2}
\int dh dm d \psi e^{\Delta t N e^{-\gamma}
   f_0(M_l) e^{B m} e^{-\psi m} -\Delta t N d_0(M_l)}
\nonumber \\ && \times e^{  \psi d -
 h(m-d)}
\nonumber \\ &=&\frac{1}{2 \pi}
\int dh dm \bigg[ \delta(d)e^{-\Delta t N d_0(M_l)} + \Delta t N e^{-\gamma}
 f_0(M_l)
\nonumber \\ && \times
 e^{B m} \delta(m-d) \bigg]
e^{-h (m-d)} + O[(\Delta t)^2]
\nonumber \\ &=&
\int dm \bigg[ \delta(d) \delta(m-d)e^{-\Delta t N d_0(M_l)}
 + \Delta t N e^{-\gamma}
f_0(M_l)
\nonumber \\ && \times
e^{B m} \delta(m-d)
\delta(m-d)
 \bigg]
\nonumber \\ &=&
\delta(d)e^{-\Delta t N d_0(M_l)}
 +  \Delta t N e^{-\gamma} f_0(M_l) e^{B d} \delta(0)
\nonumber \\ &=&
\delta(0) \left[
\Delta(d)e^{-\Delta t N d_0(M_l)}
 + \Delta t N e^{-\gamma} f_0(M_l) e^{B d}
\right] \ ,
\label{a7}
\end{eqnarray}
where $\Delta t = \beta / L$.
We note that, had we used a Fourier representation of the delta
function on the finite domain
$[-A/2, A/2]$, instead of the infinite domain $(-\infty, \infty)$,
the expression $2 \pi \delta(0)$ simply becomes $A$;
moreover such a finite representation of the delta function
is a sufficiently accurate representation
of the $\Delta(d_l)$ constraint
when $A \gg N$.
Ignoring the constant prefactor
$\delta(0)$ terms,
we can write the full partition function as
\begin{eqnarray}
Z &=& Tr \int
{\cal D}\psi
{\cal D}h
{\cal D}m
\nonumber \\ &&
\times e^{(\beta N/L) \sum_l \left[ e^{-\gamma} f_0(M_l)
e^{B m_l} e^{-\psi_l m_l} - d_0(M_l)
\right]}
\nonumber \\ && \times e^{
\sum_l \psi_l d_l
+ (\beta N/L) \sum_l \left[ -h_l m_l + h_l d_l \right]
} \ .
\label{a8}
\end{eqnarray}
We now introduce the integral representation of the constraint
$\delta[ (\beta/L) ( N M_l - \sum_n s^l_n)]$.
After rescaling $B m_l \to m_l$, $h_l \to B h_l$ we find
\begin{eqnarray}
Z &=& Tr \int
{\cal D}\psi
{\cal D}h
{\cal D}m
{\cal D}H
{\cal D}M
e^{(\beta N/L) \sum_l [e^{-\gamma} f_0(M_l) e^{m_l}  }
\nonumber \\ && \times e^{e^{-\psi_l m_l/B}
-d_0(M_l) - h_l m_l - H_l M_l] }
\nonumber \\ &&
\times e^{ (\beta / L) \sum_l H_l  \sum_n s^l_n
+ \sum_l (\psi_l + \beta N B h_l/L) \sum_n (s^{l-1}_n s^l_n -1)} \ .
\nonumber \\
\label{a9}
\end{eqnarray}
We note by an expansion of the
$\exp[(\beta N/L) e^{-\gamma} f_0(M_l) e^{m_l} \exp(-\psi_l m_l/B) ]=$\\
$\sum_{k_l=0}^\infty [(\beta N/L) e^{-\gamma} f_0(M_l) e^{m_l} ]^{k_l}
\exp(-k_l \psi_l m_l/B)/k_l!$ term
in Eq.\ (\ref{a9})
to first order in $\beta N/L$
that the integral over $\psi_l$ gives nothing more than
 $\delta(-k_l m_l/B +d_l)$
for $k_l=0,1$.  This condition, however, is already enforced by the
$h_l$ field when $k_l=1$ and by the $m_l$ field when $k_l=0$ if we
take as a rule to disallow mutations when $h_l = 0$.
We can, thus, remove the
integral over $\psi$, removing the $\delta(0)$ that we
anticipated, to find
\begin{eqnarray}
Z &=&  \int
{\cal D}h
{\cal D}m
{\cal D}H
{\cal D}M
e^{(\beta N/L) \sum_l [e^{-\gamma} f_0(M_l) e^{m_l}
-d_0(M_l)}
\nonumber \\ && \times e^{ - h_l m_l - H_l M_l ]}
 Q \ ,
\label{a10}
\end{eqnarray}
where
\begin{eqnarray}
Q = Tr
e^{(\beta /L) \sum_l  H_l  \sum_n s^l_n +
(\beta N B / L) \sum_l h_l  \sum_n (s^{l-1}_n s^l_n -1)
} F
\nonumber \\  \ .
\label{a11}
\end{eqnarray}
Here $Q$ is the partition function of $N$ 1-$d$ Ising models of length $L$.
Here $F$ enforces the constraint of disallowing mutations when  $h_l = 0$:
$F = \prod_{l=1}^L \Delta\{  \Delta(h_l) [\sum_n (s^{l-1}_n s^l_n -1)] \}$.
We note that
$Q = Q_1^N$, where $Q_1$ is the partition function of one of these
models.
We are not, at this point, allowed to assume that the $H_l$ or $h_l$ fields are
constant over $l$.  Indeed,
by Taylor series expanding the first term in Eq.\ \ref{a10} and integrating
over $m_l$, we
find that $h_l = 0$ or $h_l = L/(\beta N)$.
For $h_l = 0$, we disallow mutations, as formalized by $F$.
Thus, we can replace $e^{-2 B}$ by $\gamma t_l / N$, where $t_l = 0$
if $h_l = 0$, and $t_l = 1$ if $h_l =  L/(\beta N)$.
 We evaluate the
partition function $Q_1$ with an ordered product
of transfer matrices.  To first order
in $\beta/L$ the matrix at position $l$
is given by $T_{l} = I + \epsilon_{l}$ where
\begin{eqnarray}
 \epsilon_{l} &= &
\left(
\begin{array}{cc}
\frac{\beta H_l}{L } & \frac{\gamma t_l }{ N} \\
\frac{\gamma t_l }{ N} & -\frac{\beta H_l}{L }
\end{array}
\right)
\nonumber \\
&=&
\left(
\begin{array}{cc}
\frac{\beta H_l}{L } & \frac{\beta \gamma h_l }{ L} \\
\frac{\beta \gamma h_l }{ L} & -\frac{\beta H_l}{L }
\end{array}
\right)
 \ .
\label{a12}
\end{eqnarray}
We find
\begin{eqnarray}
Q_1 &=& Tr \prod_l T_l
\nonumber \\
&\sim&
Tr \prod_l e^{\epsilon_{l}}  \ .
 \label{a13}
\end{eqnarray}
We rescale $h \to h/\gamma$ and  $m \to m \gamma$ and
take the continuous limit to find
\begin{eqnarray}
Q_1 = Tr \hat T e^{\int_0^\beta d \beta' [ \sigma^z H(\beta') +
 \sigma^x  h(\beta')] } \ ,
 \label{a14}
\end{eqnarray}
where the operator $\hat T$ indicates (reverse) time ordering, and
$\beta' = \beta (L-l)/L$.
  We find the form of the
partition function to be
\begin{eqnarray}
Z &=& \int
{\cal D}h
{\cal D}m
{\cal D}H
{\cal D}M
\nonumber \\ &&
e^{N \int_0^\beta d \beta' [ e^{-\gamma} f_0(M) e^{\gamma m}
-d_0(M) - h m - H M]
+ N \ln Q_1 } \ .
\nonumber \\
\label{a16}
\end{eqnarray}
Noting the $N$ prefacing the entire term in the exponential, we
take the saddle point.  We note that
$\delta  Q_1 / \delta H(\beta') \vert_{H(\beta')=H, h(\beta')=h}  =
(\beta H / \sqrt{H^2 + h^2}) 2 \sinh (\beta \sqrt {H^2 + h^2})$
and \\
$\delta  Q_1 / \delta h(\beta') \vert_{H(\beta')=H, h(\beta')=h}  =
(\beta h / \sqrt{H^2 + h^2}) 2 \sinh (\beta \sqrt {H^2 + h^2})$.
We, thus, find a solution of the saddle point condition to be
fields $H, M, h, m$ independent of $\beta$ that maximize
\begin{eqnarray}
\frac{\ln Z}{N} &=& \beta \bigg[ f_0(M) e^{-\gamma} e^{\gamma m}
-d_0(M) - h m - H M]
\nonumber \\ &&
+\ln [2 \cosh (\beta \sqrt{ H^2 + h^2}) \bigg] \ ,
\label{a17}
\end{eqnarray}
when the
fields are averaged over a range $\Delta \beta = O(1/N)$ by the
saddle point limit.  In the limit of large $\beta$, we find
\begin{eqnarray}
\frac{\ln Z}{\beta N} &=& \max_{M,H,m,h} \bigg[
f_0(M) e^{-\gamma} e^{\gamma m}
-d_0(M) - h m - H M
\nonumber \\ && + ( H^2 + h^2)^{1/2} \bigg] \ .
\label{a18}
\end{eqnarray}
One can also derive Eq.\ (\ref{a18}) by means of a series
expansion in $\beta$, a ``high temperature'' expansion.

The generalization of the path integral representation
to the multiple-peak Eigen case proceeds
as in the parallel case.
One introduces $K$ fields for the magnetizations, $M^k$, and
$K$ fields enforcing the constraint, $H^k$.  One also finds in the
linear field part of the
Ising model the sum $\sum_{k=1}^K H_l^k\sum_n v^k_n s^l_n$ instead of simply
$H_l \sum_n s^l_n$.  The definition of the $y_i$ and the $\alpha_{ik}$
allows one to rewrite
this in the form that leads to Eq.\ (\ref{e7d}) or (\ref{e13}).

\bibliography{multipeak}

\begin{thebibliography}{17}
\expandafter\ifx\csname natexlab\endcsname\relax\def\natexlab#1{#1}\fi
\expandafter\ifx\csname bibnamefont\endcsname\relax
  \def\bibnamefont#1{#1}\fi
\expandafter\ifx\csname bibfnamefont\endcsname\relax
  \def\bibfnamefont#1{#1}\fi
\expandafter\ifx\csname citenamefont\endcsname\relax
  \def\citenamefont#1{#1}\fi
\expandafter\ifx\csname url\endcsname\relax
  \def\url#1{\texttt{#1}}\fi
\expandafter\ifx\csname urlprefix\endcsname\relax\def\urlprefix{URL }\fi
\providecommand{\bibinfo}[2]{#2}
\providecommand{\eprint}[2][]{\url{#2}}

\bibitem[{\citenamefont{Wright}(1931)}]{wr31}
\bibinfo{author}{\bibfnamefont{S.}~\bibnamefont{Wright}},
  \bibinfo{journal}{Genetics} \textbf{\bibinfo{volume}{16}},
  \bibinfo{pages}{97} (\bibinfo{year}{1931}).

\bibitem[{\citenamefont{Kauffmann and Levin}(1987)}]{ka87}
\bibinfo{author}{\bibfnamefont{S.}~\bibnamefont{Kauffmann}} \bibnamefont{and}
  \bibinfo{author}{\bibfnamefont{S.}~\bibnamefont{Levin}}, \bibinfo{journal}{J.
  Theor. Biol.} \textbf{\bibinfo{volume}{128}}, \bibinfo{pages}{11}
  (\bibinfo{year}{1987}).

\bibitem[{\citenamefont{Baake et~al.}(1997)\citenamefont{Baake, Baake, and
  Wagner}}]{bb97}
\bibinfo{author}{\bibfnamefont{E.}~\bibnamefont{Baake}},
  \bibinfo{author}{\bibfnamefont{M.}~\bibnamefont{Baake}}, \bibnamefont{and}
  \bibinfo{author}{\bibfnamefont{H.}~\bibnamefont{Wagner}},
  \bibinfo{journal}{Phys. Rev. Lett.} \textbf{\bibinfo{volume}{78}},
  \bibinfo{pages}{559} (\bibinfo{year}{1997}), \bibinfo{note}{{\bf 79}, 1782}.

\bibitem[{\citenamefont{Saakian and Hu}(2004{\natexlab{a}})}]{sh04a}
\bibinfo{author}{\bibfnamefont{D.~B.} \bibnamefont{Saakian}} \bibnamefont{and}
  \bibinfo{author}{\bibfnamefont{C.-K.} \bibnamefont{Hu}},
  \bibinfo{journal}{Phys. Rev. E} \textbf{\bibinfo{volume}{69}},
  \bibinfo{pages}{021913} (\bibinfo{year}{2004}{\natexlab{a}}).

\bibitem[{\citenamefont{Saakian and Hu}(2004{\natexlab{b}})}]{sh04b}
\bibinfo{author}{\bibfnamefont{D.~B.} \bibnamefont{Saakian}} \bibnamefont{and}
  \bibinfo{author}{\bibfnamefont{C.-K.} \bibnamefont{Hu}},
  \bibinfo{journal}{Phys. Rev. E} \textbf{\bibinfo{volume}{69}},
  \bibinfo{pages}{046121} (\bibinfo{year}{2004}{\natexlab{b}}).

\bibitem[{\citenamefont{Eigen}(1971)}]{ei71}
\bibinfo{author}{\bibfnamefont{M.}~\bibnamefont{Eigen}},
  \bibinfo{journal}{Naturwissenschaften} \textbf{\bibinfo{volume}{58}},
  \bibinfo{pages}{465} (\bibinfo{year}{1971}).

\bibitem[{\citenamefont{Eigen et~al.}(1989)\citenamefont{Eigen, McCaskill, and
  Schuster}}]{ei89}
\bibinfo{author}{\bibfnamefont{M.}~\bibnamefont{Eigen}},
  \bibinfo{author}{\bibfnamefont{J.}~\bibnamefont{McCaskill}},
  \bibnamefont{and} \bibinfo{author}{\bibfnamefont{P.}~\bibnamefont{Schuster}},
  \bibinfo{journal}{Adv. Chem. Phys.} \textbf{\bibinfo{volume}{75}},
  \bibinfo{pages}{149} (\bibinfo{year}{1989}).

\bibitem[{\citenamefont{Crow and Kimura}(1970)}]{ck70}
\bibinfo{author}{\bibfnamefont{J.~F.} \bibnamefont{Crow}} \bibnamefont{and}
  \bibinfo{author}{\bibfnamefont{M.}~\bibnamefont{Kimura}},
  \emph{\bibinfo{title}{An Introduction to Population Genetics Theory}}
  (\bibinfo{publisher}{Harper and Row}, \bibinfo{address}{New York},
  \bibinfo{year}{1970}).

\bibitem[{\citenamefont{Messer et~al.}(2005)\citenamefont{Messer, Arndt, and
  L{\"a}ssig}}]{Lassig}
\bibinfo{author}{\bibfnamefont{P.~W.} \bibnamefont{Messer}},
  \bibinfo{author}{\bibfnamefont{P.~F.} \bibnamefont{Arndt}}, \bibnamefont{and}
  \bibinfo{author}{\bibfnamefont{M.}~\bibnamefont{L{\"a}ssig}},
  \bibinfo{journal}{Phys. Rev. Lett.} \textbf{\bibinfo{volume}{94}},
  \bibinfo{pages}{138103} (\bibinfo{year}{2005}).

\bibitem[{\citenamefont{Saakian et~al.}(2004)\citenamefont{Saakian, Hu, and
  Khachatryan}}]{sh04c}
\bibinfo{author}{\bibfnamefont{D.~B.} \bibnamefont{Saakian}},
  \bibinfo{author}{\bibfnamefont{C.-K.} \bibnamefont{Hu}}, \bibnamefont{and}
  \bibinfo{author}{\bibfnamefont{H.}~\bibnamefont{Khachatryan}},
  \bibinfo{journal}{Phys. Rev. E} \textbf{\bibinfo{volume}{70}},
  \bibinfo{pages}{041908} (\bibinfo{year}{2004}).

\bibitem[{\citenamefont{Jain and Krug}(2005)}]{Krug}
\bibinfo{author}{\bibfnamefont{K.}~\bibnamefont{Jain}} \bibnamefont{and}
  \bibinfo{author}{\bibfnamefont{J.}~\bibnamefont{Krug}}, in
  \emph{\bibinfo{booktitle}{Structural approaches to sequence evolution:
  Molecules, networks and populations}}, edited by
  \bibinfo{editor}{\bibfnamefont{H.~R.} \bibnamefont{U.~Bastolla},
  \bibfnamefont{M.~Porto}} \bibnamefont{and}
  \bibinfo{editor}{\bibfnamefont{M.}~\bibnamefont{Vendruscolo}}
  (\bibinfo{publisher}{Springer Verlag}, \bibinfo{address}{Berlin},
  \bibinfo{year}{2005}), \bibinfo{note}{q-bio.PE/0508008}.

\bibitem[{\citenamefont{Peliti}(2002)}]{Peliti2002}
\bibinfo{author}{\bibfnamefont{L.}~\bibnamefont{Peliti}},
  \bibinfo{journal}{Europhys. Lett.} \textbf{\bibinfo{volume}{57}},
  \bibinfo{pages}{745} (\bibinfo{year}{2002}).

\bibitem[{\citenamefont{Deem and Lee}(2003)}]{de03}
\bibinfo{author}{\bibfnamefont{M.~W.} \bibnamefont{Deem}} \bibnamefont{and}
  \bibinfo{author}{\bibfnamefont{H.~Y.} \bibnamefont{Lee}},
  \bibinfo{journal}{Phys. Rev. Lett.} \textbf{\bibinfo{volume}{91}},
  \bibinfo{pages}{068101} (\bibinfo{year}{2003}).

\bibitem[{\citenamefont{Tarazona}(1992)}]{Tarazona}
\bibinfo{author}{\bibfnamefont{P.}~\bibnamefont{Tarazona}},
  \bibinfo{journal}{Phys. Rev. A} \textbf{\bibinfo{volume}{45}},
  \bibinfo{pages}{6038} (\bibinfo{year}{1992}).

\bibitem[{\citenamefont{Baake et~al.}(1998)\citenamefont{Baake, Baake, and
  Wagner}}]{bb98}
\bibinfo{author}{\bibfnamefont{E.}~\bibnamefont{Baake}},
  \bibinfo{author}{\bibfnamefont{M.}~\bibnamefont{Baake}}, \bibnamefont{and}
  \bibinfo{author}{\bibfnamefont{H.}~\bibnamefont{Wagner}},
  \bibinfo{journal}{Phys. Rev. E} \textbf{\bibinfo{volume}{57}},
  \bibinfo{pages}{1191} (\bibinfo{year}{1998}).

\bibitem[{\citenamefont{Eigen}(2002)}]{ei02}
\bibinfo{author}{\bibfnamefont{M.}~\bibnamefont{Eigen}},
  \bibinfo{journal}{Proc. Natl. Acad. Sci. USA} \textbf{\bibinfo{volume}{99}},
  \bibinfo{pages}{13374} (\bibinfo{year}{2002}).

\bibitem[{\citenamefont{Park and Deem}(2004)}]{de04}
\bibinfo{author}{\bibfnamefont{J.~M.} \bibnamefont{Park}} \bibnamefont{and}
  \bibinfo{author}{\bibfnamefont{M.~W.} \bibnamefont{Deem}},
  \bibinfo{journal}{Physica A} \textbf{\bibinfo{volume}{341}},
  \bibinfo{pages}{455} (\bibinfo{year}{2004}).

\end{thebibliography}

\end{document}